# MAC Hurdles in Body Sensor Networks

Sana Ullah[1], Pervez Khan[1], Young-Woo Choi[2], Hyung-Soo Lee[2], and Kyung Sup Kwak[1]

[1]Graduate School of IT and Telecommunications, Inha University
253 Yonghyun-Dong, Nam-Gu, Incheon 402-751, South Korea.
E-mail. sanajcs@hotmail.com, pervaizkanju@hotmail.com, kskwak@inha.ac.kr

[2]Electronics and Telecommunication Research Institute
1104-302 YeolMae Apt., Noeun-Dong, Yusung-Gu, Daejeon, South Korea
ywchoi@etri.re.kr , hsulee@etri.re.kr

***Abstract* — The last few decades have seen considerable research progress in microelectronics and integrated circuits, system-on-chip design, wireless communication, and sensor technology. This progress has enabled the seamless integration of autonomous wireless sensor nodes around a human body to create a Body Sensor Network (BSN). The development of a proactive and ambulatory BSN induces a number of enormous issues and challenges. This paper presents the technical hurdles during the design and implementation of a low-power Medium Access Control (MAC) protocol for in-body and on-body sensor networks. We analyze the performance of IEEE 802.15.4 protocol for the on-body sensor network. We also provide a comprehensive insight into the heterogeneous characteristics of the in-body sensor network. A low-power technique called Pattern-Based Wake-up Table is proposed to handle the normal traffic in a BSN. The proposed technique provides a reliable solution towards low-power communication in the in-body sensor network.**

## 1. Introduction

The remote monitoring of body status, and the surrounding environment, are becoming more important for sporting activities, the safety of members of the emergency services, members of the military and health care. The levels of fitness required for the very competitive international sporting events require athletes to be at the very pinnacle of fitness with every muscle used to its utmost. Furthermore, many body functions are traditionally monitored only rarely and separated by a considerable period of time. This can give a very incomplete picture of what is really happening. Consider a patient visiting a doctor for a blood pressure check; he/she may be anxious and thus have elevated pressure resulting in an inaccurate diagnosis. If, however, the patient can be fitted with a simple monitoring system that requires no intervention, then a picture can be built up of how the pressure changes through the day when he/she goes about their normal business. This will give a better picture of what is happening and remove inaccurate results caused by going to visit the doctor. To achieve these requirements, monitoring of movement and body function are essential. This monitoring requires the sensors and wireless system to be very lightweight and to be integrated un-obtrusively into the clothing.

A Body Sensor Network (BSN) allows the integration of intelligent, miniaturized, low power, invasive and non-invasive sensor nodes to monitor body function and the surrounding environment. Each intelligent node has enough capability to process and forward information to a base station for diagnosis and prescription. A BSN provides long term health monitoring of patients under natural physiological states without constraining their normal activities. It can be used to develop a smart and affordable health care system and can be a part of diagnostic procedure, maintenance of chronic condition, supervised recovery from a surgical procedure and to handle emergency events [1]. Monitoring emergency services and the military both have similar requirements. They both may have individuals out of sight and in danger such that rescue is needed. They both need easy to use devices that will reliably transmit data without getting in the way of the prime function. The scope of a BSN spans around three domains; Off-body communication, On-body communication and In-body communication. Off-body communication is the communication from the base station to the transceiver on human side. On-body communication is the communication with on-body networks and wearable system. In-body communication is the communication between invasive or implantable devices and external monitoring equipment. Monitoring in-body functions and the ability to communicate with an implanted therapeutic device, such as a pacemaker, are essential for its best use. Applications include monitoring and programme changes for pacemakers and implantable cardiac defibrillators, control of bladder function and restoration of limb movement. The space within a body is very limited and the available materials are few. These applications may require continuous or occasional one or two-way transmission. Some applications will require a battery where the current drain must be low so as not to reduce the working life of the implant function. If the data transfer is small and only occasional then an “RFID” technique can be employed. This is relevant for an intracranial pressure monitor.

The development of an affordable BSN induces a number of issues and challenges such as interoperability, scalability, Quality of Service (QoS), and energy efficient communication. There are various low-power techniques to ensure energy efficient communication in a wireless sensor network such as fixed duty cycling technique in SMAC [2] and wake-up slots in TDMAW [3]. However, they are not energy efficient in case of a heterogeneous BSN. In this paper, we discuss the technical hurdles in the development of an energy efficient low-power MAC protocol for in-body and on-body sensor networks.

The following paper is divided into four sections. Section 2 presents a general discussion on MAC protocol for a BSN. Section 3 and 4 present a brief analysis on in-body and on-body MAC protocols. The final section concludes our work.

## 2. BSN-MAC

Unlike SMAC and TMAC [4], the traffic characteristics in a Body Sensor Network (BSN) vary from periodic to non-periodic and vice versa. The concept of fixed duty cycling technique gives limited answer when it comes to the heterogeneous behavior of autonomous sensor nodes in a BSN. The dynamic nature of these nodes does not urge synchronized periodic wakeup periods. Some nodes, e.g., electrocardiogram (ECG), may send data at 1/hour rate to the coordinator, while other may send data twice in a week. These nodes should also have the capabilities to sense and transmit emergency information. The data is classified into three categories, i.e., Normal Traffic, On-Demand Traffic, and Emergency Traffic. Figure 1 shows the MAC mapping of classified data before transmission.

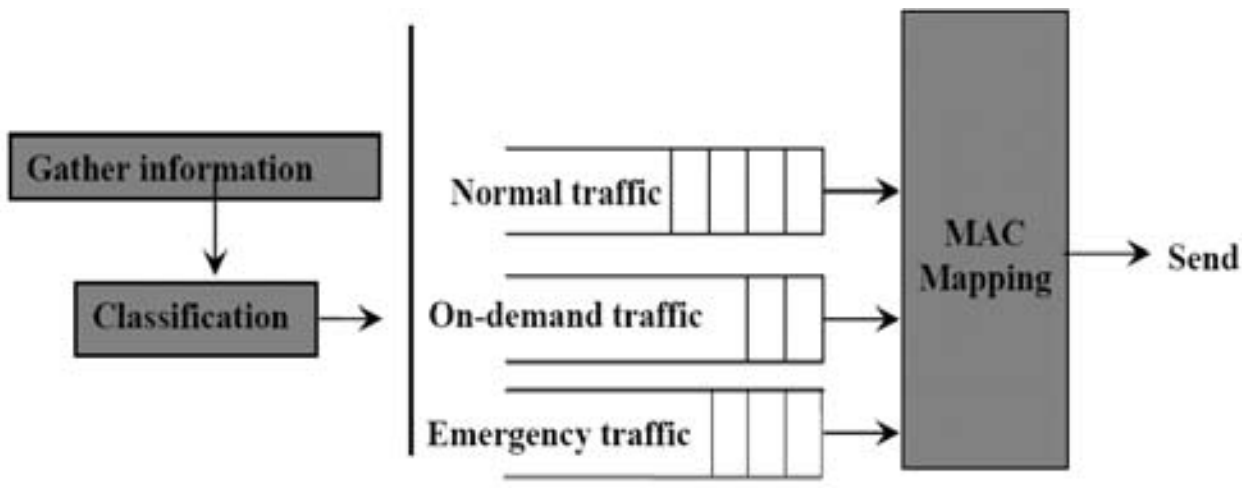

**Figure 1. Traffic Classification**

The BSN MAC protocol is required to accommodate the entire traffic classification in a sophisticated manner. Most of the well-known low-power MAC protocols such as IEEE 802.15.4 [5], SMAC, TMAC, and WiseMAC [6] cannot accommodate these diverse traffic requirements. They give limited answers in terms of energy efficiency and reliability. Furthermore, they cannot handle both medical and non-medical BSN applications. Medical data usually needs high priority and reliability than non-medical data. Time critical event needs highest reliability. IEEE 802.15.4 Guaranteed Time Slot (GTS) can be utilized to handle time critical events but they expire in case of low traffic.

The IEEE 802.15.6 aims to provide low-power in-body and on-body wireless communication standard for medical and non-medical applications [7]. On-body sensor network has an immediate market demand and is easy to standardize without considering the in-body communication. The committee has suggested four options to design MAC and PHY layer for a BSN.

1- To define MAC and PHY standard for on-body communication to serve immediate market needs. A slight modification to the existing MAC standard with an alternative PHY layer for in-body communication is also suggested.

2- To define MAC and PHY only for on-body communication to serve immediate market needs.

3- To define MAC and PHY only for in-body communication.

4- To define MAC and PHY for in-body and on-body communication simultaneously regardless of their affects on the availability of specification.

Splitting the task group into two groups only for multiple PHY layer is not an attractive approach. We can define two different PHY's in one standard addressing different requirement as illustrated in Figure 2. However, in-body and on-body communications have different lead times to standard completion.

| **BSN MAC** | |
|---|---|
| **MICS Band** | **ISM/UWB** |

**Figure 2. One MAC and Multi PHY**

## 3. On-Body MAC

On-body sensor networks comprise of non-invasive sensor nodes used for various applications, ranging from medical to interactive gaming and entertainment applications. They use Industrial, Scientific, and Medical (ISM) or Ultra Wide Band (UWB) band for data transmission. A performance analysis of IEEE 802.15.4 standard has been presented in [8], where a star network configuration with a non-beacon enabled mode has been adapted to prolong lifetime of nodes from 10 to 15 years. However, in a non-beacon enabled mode, the coordinator cannot communicate with the nodes until invited by them. In a beacon-enabled mode, the coordinator can initiate the communication with the nodes but all nodes must wake up to receive the beacon frames, which is not suitable for a BSN where some nodes do not always send the data.

We analyze the performance of IEEE 802.15.4 (beacon-enabled mode), Preamble-Based TDMA [9], and SMAC protocols for on-body sensor networks. Our analysis is verified by extensive simulation using NS-2[10]. Simulation results show that IEEE 802.15.4, when configured in a beacon-enabled mode, outperforms SMAC and PB-TDMA as shown in Figure 3. However, the precise location of nodes and the body position influence the packet delivery ratio.

Intel Corporation conducted a series of experiments to analyze the performance of IEEE 802.15.4 for an on-body sensor network [11]. They deployed a number of Intel Mote 2 [12] devices on chest, waist, and the right ankle. Figure 4 shows the packet delivery ratio at 0dBm transmit power when a person is standing. The connection between ankle and waist cannot be established, even for a short distance of 1.5m. All other connections show favourable performance. A modification of IEEE 802.15.4 standard towards on-body sensor networks could give reasonable solution for some of the medical applications, but limits its performance for non-medical applications.

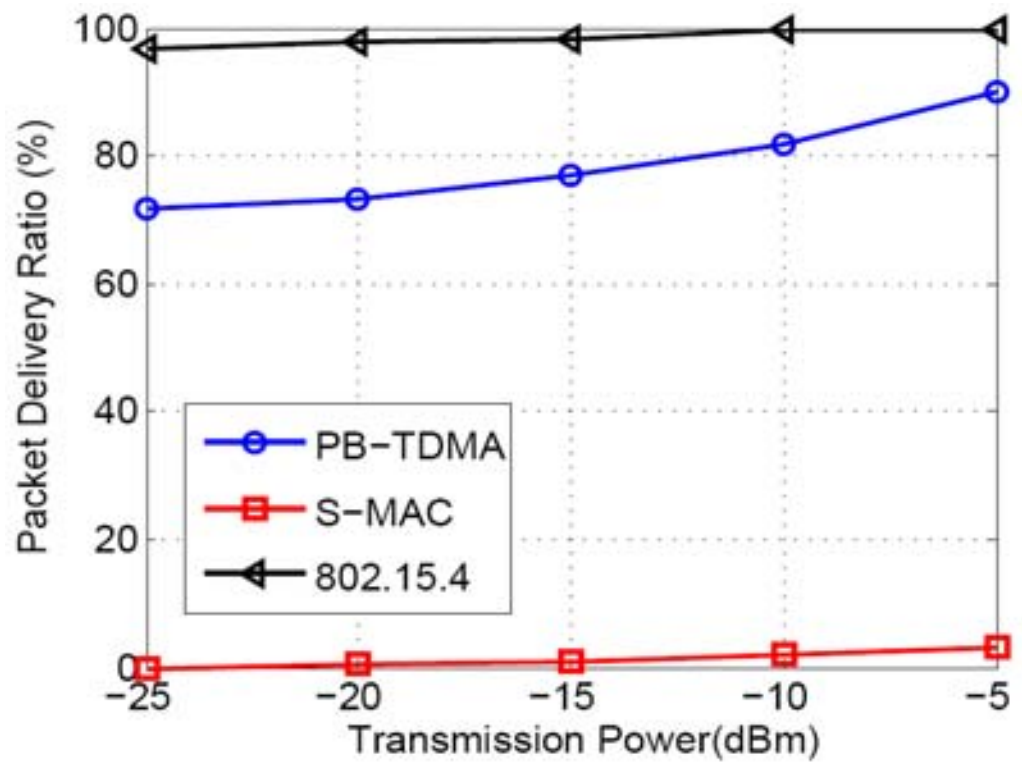

**Figure 3. Packet Delivery Ratio**

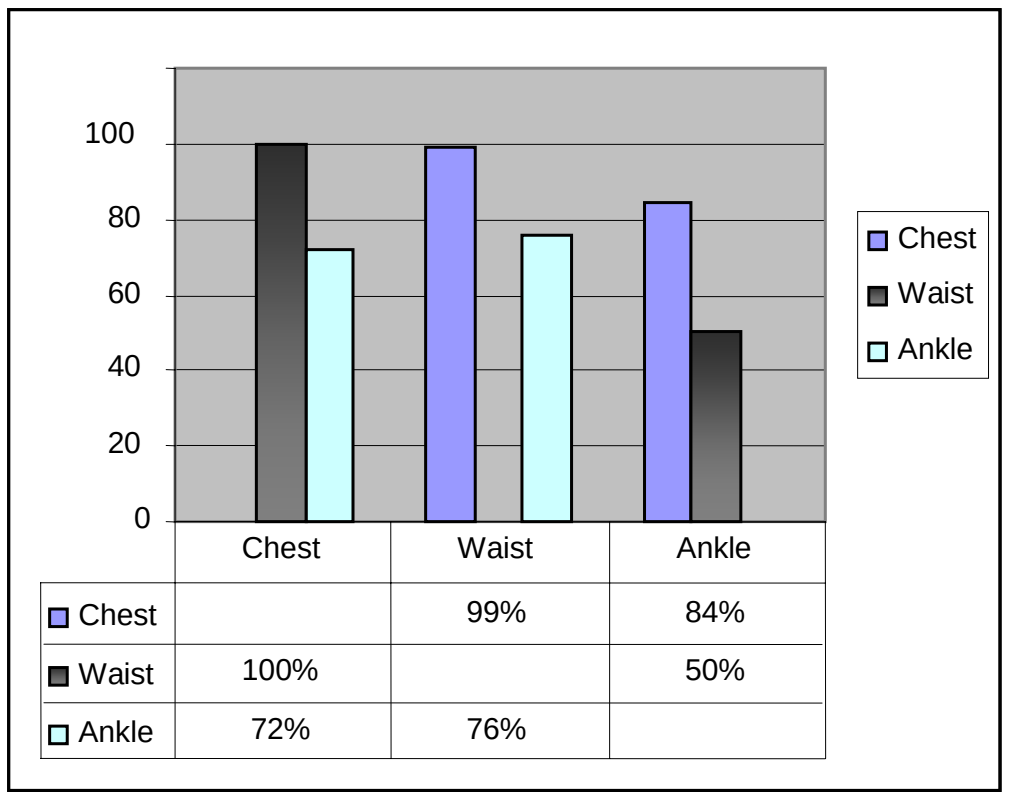

**Figure 4. Packet Delivery Ratio of IEEE 802.15.4 when a person is standing**

## 4. In-Body MAC

The most challenging task in developing a low-power BSN-MAC protocol is to accommodate in-body sensor nodes in a more efficient way. In-body nodes are implanted under human skin and have critical power requirements. They are totally different than on-body nodes in terms of power efficiency and data transmission rate (10kbps for medical and up to 10Mbps for non-medical applications). Moreover, they need to send emergency data in less than 1 second to the coordinator. This is a hot issue in the design and implementation of an in-body MAC. The nodes are required to be self triggered when exceeds a predefine threshold for emergency situation. Critical data requires low latency and high reliability. The solution is to adjust initial back-off windows for critical and non-critical traffics [13]. Non-critical traffic nodes have larger initial back-off window than critical traffic nodes, i.e., ($W_0^{\alpha} \leq W_0^{\beta}$) where $W_0^{\alpha}$ is initial back-off window for critical traffic nodes $\alpha$ and $W_0^{\beta}$ is initial back-off window for non-critical traffic nodes $\beta$. Figure 5 illustrates the average packet latency for a number of in-body nodes. The smaller initial back-off window for the critical nodes results in lower latency. The figure also shows packet latency of standard CSMA/CA scheme used in IEEE 802.15.4.

The use of CSMA/CA for in-body communication does not provide reliable solution [14]. The main reason is that the path loss inside human body due to tissue heating is much higher than the path loss in free space. The in-body nodes cannot perform Clear Channel Assessment (CCA) in a favorable way. Figure 6 shows the CCA for in-body and on-body nodes. For a given -85dBm CCA threshold, the on-body nodes cannot see the activity of in-body nodes when they are away at 3m from the surface of the body.

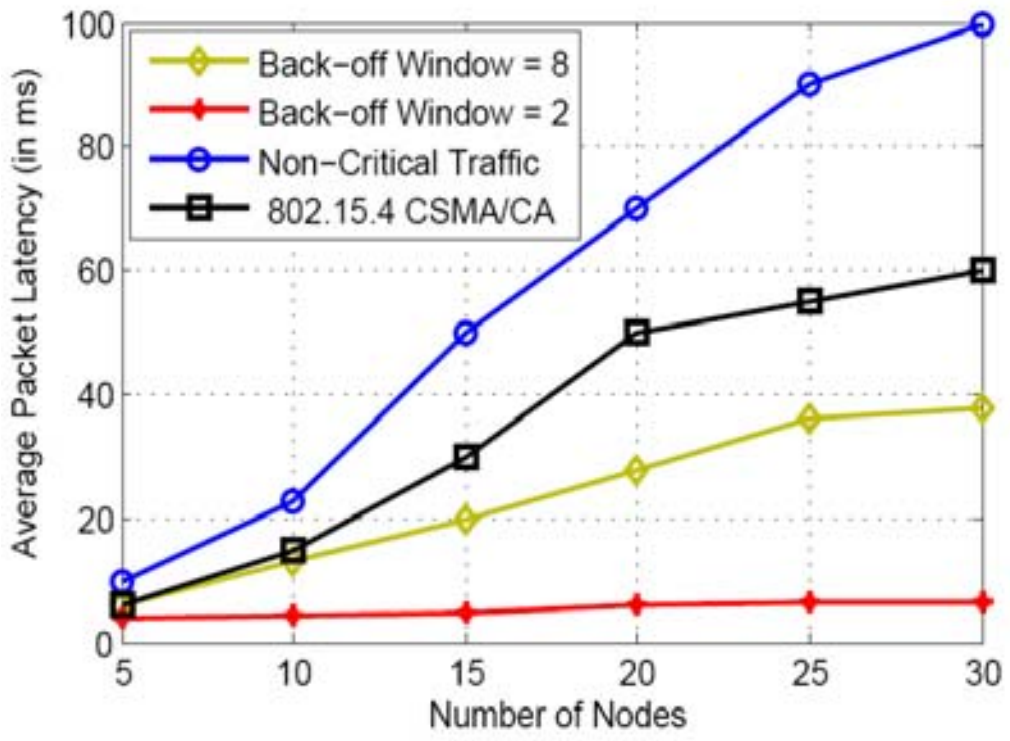

**Figure 5. Average Packet Latency**

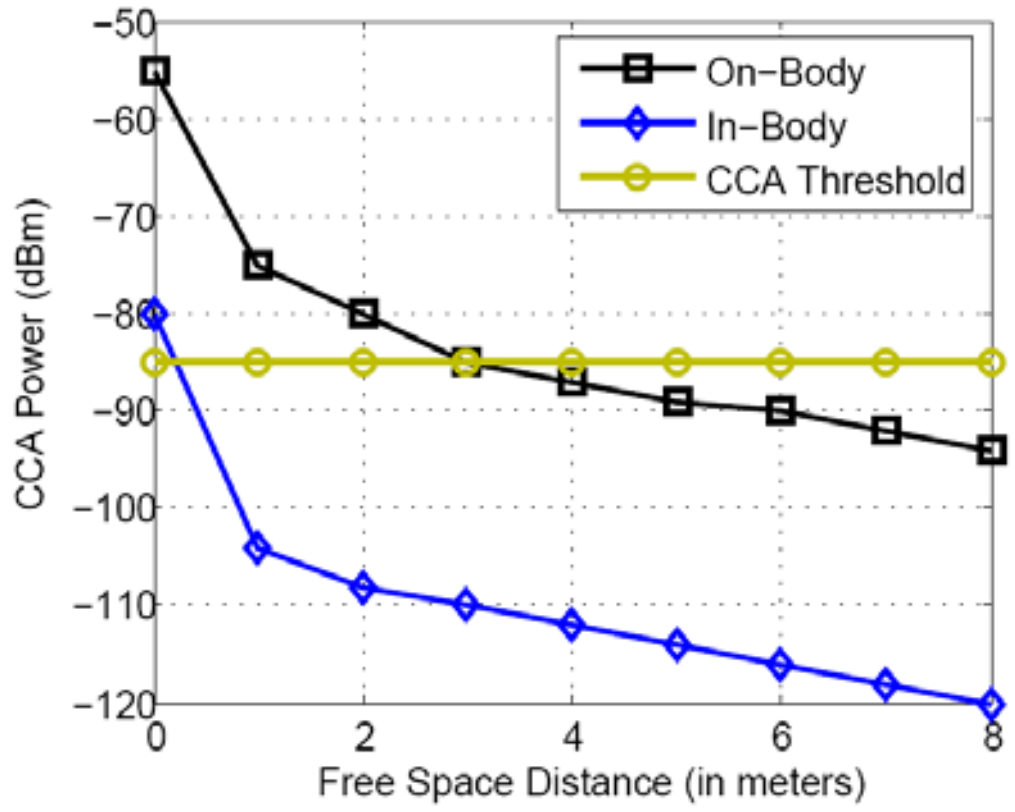

**Figure 6. CCA for In-body and On-body nodes**

The alternative approach is to use a TDMA-Based solution to accommodate the entire traffic classification of in-body nodes. Table 1 shows the traffic classification and the corresponding solutions. The on-demand traffic is further divided into continuous and non-continuous traffics. The doctor may need continuous flow of information in case of a surgery, or some images for diagnosis and prescription. We present a solution to handle the normal traffic in the in-body sensor network. The communication is based on pre-defined (by company) or later modified (by doctor) wake-up patterns that indicate the traffic nature of the nodes. The wake-up patterns of all nodes are organized into a table called Pattern-Based wake-up table. The wake-up table is maintained by the coordinator/doctor. Based on the node's wake-up pattern, the coordinator can calculate its own wake-up traffic. This could save energy of the coordinator. The coordinator does not need to stay active when there is no traffic from the nodes. Table 2 is an example of a Pattern-Based Wake-up Table maintained by the coordinator.

**Table 1. Traffic Classification in the In-body Sensor Network**

| Devices | Normal Traffic | Emergency Traffic | On-demand traffic | |
|---|---|---|---|---|
| | | | Continuous | Non-continuous |
| Sensor nodes | Send data based on wake-up patterns | Sensor are triggered when exceeds a predefined threshold and Send a Wake-up Signal (Wake-up Radio) | Receives wake-up signal from coordinator in case of e.g., Surgical events | Receives wake-up signal from coordinator when requires limited information (not continuous) e.g., images |
| Coordinator | Receive data | Receive Wake-up Signal | Send a wake-up signal (Wake-up Radio) | Send a wake-up signal (Wake-up Radio) |

**Table 2. Pattern-Based Wake-up Table**

| In-body sensor nodes | Wake-up Patterns | | |
|---|---|---|---|
| Blood Pressure (BP) | 0 | 1 | 1 |
| Electrocardiogram (ECG) | 1 | 0 | 0 |
| Electromyogram (EMG) | 0 | 1 | 0 |
| BSN Coordinator | 1 | 1 | 1 |

In the above table, each node has a wake-up pattern represented by 0 and 1. The digit 0 keeps the nodes into sleep mode while 1 keeps them active for transmission. For example, the wake-up pattern of BP node (011) means that it needs to sleep during the first slot duration indicated by 0 and transmits data during second and third slot duration indicated by 1. The coordinator calculates its own wake-up pattern (111) based on the node's pattern. The coordinator goes into sleep mode when all nodes do not send data, i.e., when their wake-up patterns contain only 0. The number of 1's in the same column does not indicate traffic at the same time *t*. The wake-up table is based on TDMA scheme. Figure 7 represents the above table in a set of TDMA frames.

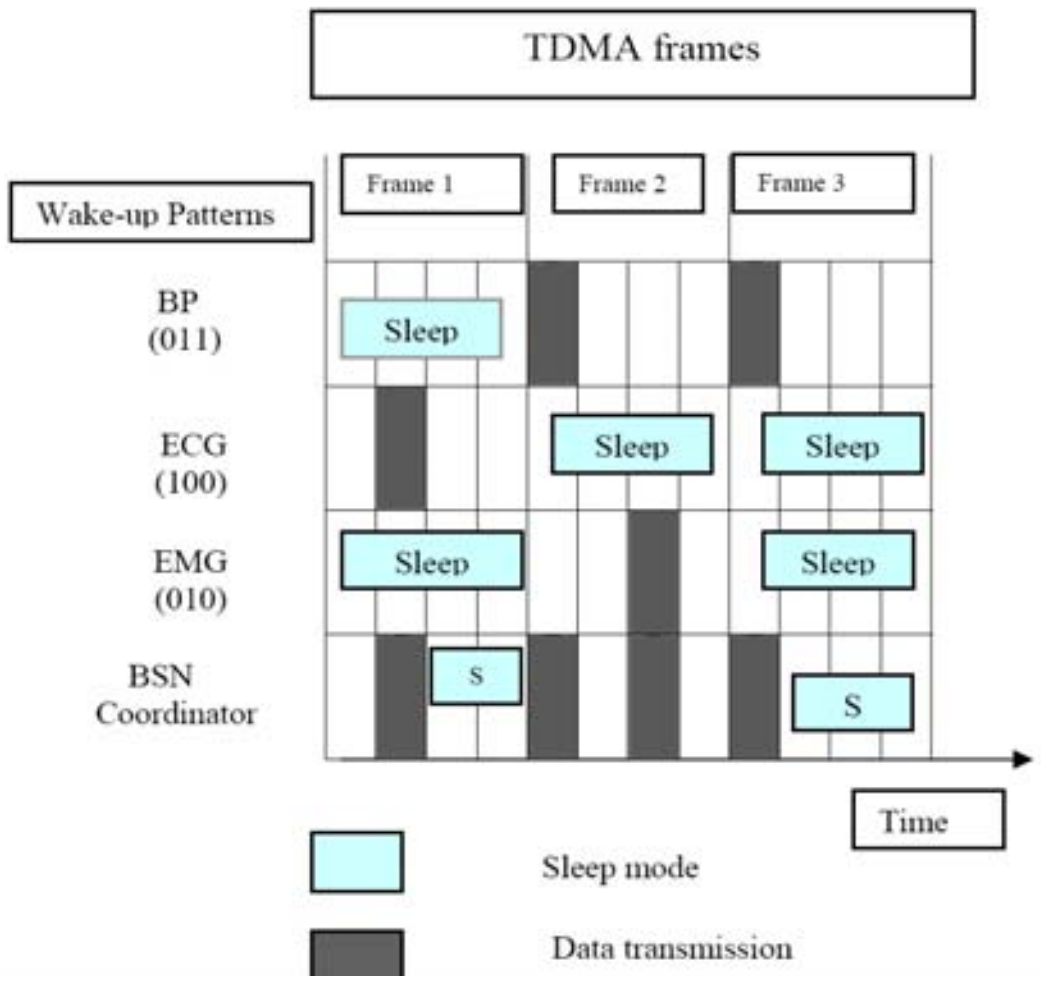

**Figure 7. Pattern-Based TDMA**

Each node transmits data in its own TDMA slot. For example, a BP node transmits in the first, an ECG in the second, and an EMG in the third TDMA slot. For the wake-up pattern 011, the BP node sleeps during the first TDMA frame and transmits data to the coordinator in the first slots of second and third TDMA frames.

The pattern-based wake-up table is used to handle normal traffic in the in-body sensor network. For emergency and on-demand traffics, the nodes and the coordinator send a wake-up signal of a very short duration. However, traditional wake-up radio concepts have several limitations when considered in the in-body sensor network. They are not able to wake-up a particular node. All nodes wake-up in response to a single wake-up signal, which is not the required environment. The use of different radio frequencies to wake-up a particular node may provide an optimal solution.

## 5. Conclusions

The traffic in a BSN is classified into normal, emergency, and on-demand traffics. None of the existing MAC protocols can accommodate these diverse traffic requirements in a power efficient manner. We presented a performance analysis of various MAC protocols including IEEE 802.15.4 for on-body sensor networks. The on-body sensor networks comprise of non-invasive sensor nodes used for medical, interactive gaming and entertainment applications. They are totally different than in-body sensor networks in terms of power consumption and data transmission rate. In-body sensor nodes are implanted under human skin and require efficient power management techniques to prolong sensor life time up to many years. The use of CSMA/CA is not reliable due to high path loss inside the human body. We suggested a TDMA based solution to handle normal traffic in the in-body sensor networks. Our proposed method provided a reliable solution towards energy efficient communication in the in-body sensor networks.

## Acknowledgement

This work was supported by the IT R&D program of MKE/IITA, [2008-F-050-01, Development of WBAN system for In-body and On-body]